\documentclass[letterpaper,twocolumn,english,showpacs,preprintnumbers,amsmath,amssymb]{revtex4-1}
\usepackage{float}
\usepackage{amsmath}
\usepackage{amssymb}
\usepackage{graphicx}
\usepackage{esint}

\makeatletter

 \@ifundefined{textcolor}{}
 {%
   \definecolor{BLACK}{gray}{0}
   \definecolor{WHITE}{gray}{1}
   \definecolor{RED}{rgb}{1,0,0}
   \definecolor{GREEN}{rgb}{0,1,0}
   \definecolor{BLUE}{rgb}{0,0,1}
   \definecolor{CYAN}{cmyk}{1,0,0,0}
   \definecolor{MAGENTA}{cmyk}{0,1,0,0}
   \definecolor{YELLOW}{cmyk}{0,0,1,0}
 }

\usepackage{color}
\usepackage{epsfig}%
\usepackage{subfigure}%
\usepackage{dcolumn}%
\usepackage{stmaryrd}%
\usepackage{mathrsfs}%
\usepackage{pifont}%
\usepackage{amsthm}%
\usepackage{bm}%
\usepackage{latexsym}%
\usepackage{amsfonts}%
\newcommand{\beq}{\begin{equation}}
\newcommand{\eeq}{\end{equation}}
\newcommand{\beqa}{\begin{eqnarray}}
\newcommand{\eeqa}{\end{eqnarray}}

\makeatother

\usepackage{babel}
\begin{document}
%*******************************************************************************
%                                    TITLE
\title{Restoration of Quantum State in Dephasing Channel}
%*******************************************************************************
%*******************************************************************************
%                                   BYLINES
\author{Xinyu Zhao}\email[]{xzhao1@stevens.edu}
\author{Samuel R. Hedemann}
\author{Ting Yu}\email[]{Ting.Yu@stevens.edu}
\affiliation{Center for Controlled Quantum Systems and the Department of Physics
and Engineering Physics, Stevens Institute of Technology, Hoboken,
New Jersey 07030, USA}
\date{\today}
%*******************************************************************************
%*******************************************************************************
\begin{abstract}%                 0. ABSTRACT
In this paper, we propose an explicit scheme to fully recover a multiple-qubit
state subject to a phase damping noise. We establish the theoretical
framework and the operational procedure to restore an unknown initial
quantum state for an $N$-qubit model interacting with either individual
baths or a common bath. We give an explicit construction
of the \textit{random unitary} (RU) Kraus decomposition for an $N$-qubit
model interacting with a common bath. We also demonstrate how to use only
one unitary reversal operation to restore an arbitrary state with
phase damping noise. In principle, the initial state can always be
recovered with a success probability of $1$. Interestingly, we found that non-RU decomposition can also be used to restore some
particular entangled states. This may open a new path to restore a
quantum state beyond the standard RU scheme.
\end{abstract}
%*******************************************************************************
%*******************************************************************************
%                             PACS & TITLE COMMAND
%The definitions of these should be written here.
\maketitle
%*******************************************************************************
%*******************************************************************************
%                              I. INTRODUCTION
\section{\label{sec:I}Introduction}
Recent research has focused on protection or restoration of a quantum
state subject to the influence of intrinsic and extrinsic
noises \cite{Book1,Book2,QCQI}. Quantum noise is ubiquitous and is
typically registered as disentanglement and loss of quantum coherence
\cite{WW,XX,YY,Yu-Eberly04}. Several theoretical schemes have
been proposed to control decoherence, ranging from dynamic decoupling
control \cite{DD} and feedback control \cite{FBC}, to weak measurement
\cite{WK1,WK2,WK3} and error correction codes \cite{ECC}, etc. While
the dynamic decoupling and weak measurement are implemented to protect
the quantum states by active external interventions, the error correction
in quantum codes can be achieved through a passive realization of
the encoded systems. However, in all realistic applications, the errors
in quantum operations utilized in quantum information processing cannot
be completely eliminated.  Besides, the intricacy in manipulating coupled
multiple-qubit systems has posed a serious challenge in scalable realizations
in various quantum information carriers. All these fundamental issues
of decoherence and control are still active research subjects.

Quantum decoherence can be modeled as a non-unitary process, and the
non-unitary dynamics of a quantum open system can be described by a master
equation. Clearly, a state evolving unitarily in an ideally \textit{closed}
system can be recovered by one unitary reversal operation. Therefore
it is natural to ask, \textit{is it possible to recover an initial
quantum state for a non-unitary evolution via a unitary quantum reversal
operation?}

The state of an quantum open system, denoted by the reduced density
operator $\rho_{S}$ at time $t$, may be written in the form of Kraus
(operator sum) representation \cite{Choi,Lind,Kraus,QCQI}: $\rho_{S}(t)={\textstyle \sum_{n}}K_{n}(t)\rho_{S}(0)K_{n}^{\dagger}(t)$
with $\sum_{n}K_{n}^{\dagger}K_{n}=I$. If these Kraus operators $K_{n}$
are proportional to a set of unitary operators as $K_{n}=c_{n}U_{n}$,
then restoration of the initial quantum state is possible \cite{LostFound},
and such a decomposition is called a \textit{random
unitary} (RU) decomposition with RU-type Kraus operators $\{K_{n}\}$. In a well-defined situation where the
environment is measurable, M. Gregoratti and R. F. Werner explicitly
provide an error correction scheme which needs only one reversal operation
on the system based on the outcome of a measurement
performed on the environment \cite{LostFound}. Ideally, the success
probability is always $1$, and the fidelity of the recovered state
is also $1$. This restoration scheme is of interest in many scenarios
where a quantum measurement may be performed on the environment 
under consideration. However,  generally it is not clear whether or not the 
required RU decomposition
always exists. In fact, it has been shown that the RU decomposition does
not exist in a generic dissipative model \cite{EC_dissipative}. 
Even in cases for which it \textit{does} exist, finding the RU-type Kraus operators is often difficult. For the dephasing channel, Buscemi \textit{et al.}~prove
that if the dimension $d$ of the Hilbert space satisfies $d\leqslant3$,
then the RU decomposition \textit{does} always exist \cite{Buscemi2}. Moreover,
Strunz and co-workers have provided an explicit example of the RU
decomposition for the one-qubit dephasing channel \cite{Strunz}.
However, the explicit construction of an RU decomposition for an $N$-qubit
system coupled to a general environment has not yet been proposed.

The purpose of this paper is to explicitly construct an RU decomposition
for an $N$-qubit model in both the individual baths case and the
common bath case. We investigate the feasibility of performing environment-assisted error correction for dephasing chanel. 
Since a major limitation of the original scheme \cite{LostFound} is that RU decomposition
rarely exists for certain models, we provide an example of how to use non-RU decomposition to fully recover some entangled states, 
and thus overcome the RU limitation.  Moreover, in our scheme, the measurement performed on the environment is in the Fock basis,
the restoration, in many interesting cases, only depends on the parity of total photon numbers (even or odd) in the environment. 
This may enhance the feasibility of experimental realizations of the proposed restoration scheme.

This paper is organized as follows: in Sec.~\ref{sec:II}, we review
the general error correction scheme proposed in \cite{LostFound}
and developed in \cite{EC_Depolarizing,EC_dissipative,Buscemi1,Buscemi2,Strunz,Audenaert2008,Rosgen2008,Hayden2005,Lidar2010,Exp_ECC}.
In Sec.~\ref{sec:III}, we investigate the case of an $N$-qubit
model interacting with a common bath. Particularly, we explicitly
construct the RU decomposition and non-RU decomposition for this $N$-qubit
system. When the RU decomposition exists, the initial state may be
recovered using the standard restoration scheme. Interestingly, our work goes beyond the standard restoration
scheme based on the RU decomposition \cite{LostFound}. We show that,
for a non-RU decomposition, a quantum operation can entail a full
recovery of some entangled states embedded in the phase noise. In
Sec.~\ref{sec:IV}, we investigate another type of interaction between
system and baths, $N$ qubits interacting with individual baths. We
first derive the RU-type Kraus decomposition for a single-qubit system
associated with an explicit measurement basis. Then, we extend the
restoration scheme to the $N$-qubit model interacting with individual
dephasing baths. Thus, we show that perfect restoration can always
be achieved for a multiple-qubit system for both the common bath case
and the individual baths case.
%                                  END of I
%*******************************************************************************
%*******************************************************************************
%            II. Error Correction Scheme Based on RU Decomposition
\section{\label{sec:II}Error Correction Scheme Based on RU Decomposition}
The framework of open quantum systems is established from quantum
theory for the total system (system plus environment), which is governed
by the Schr\"{o}dinger equation. The evolution of the total density
matrix is given by $\rho_{{\rm tot}}(t)=U\rho(0)U^{\dagger}$, where
$U=e^{-iH_{{\rm tot}}t}$ (setting $\hbar=1$) is the total evolution
operator. The evolution for the reduced density matrix of the system
can be obtained by tracing over the environmental degrees of freedom
$\rho_{S}(t)=\text{tr}_{E}[\rho_{{\rm tot}}(t)]$, and it can formally
be written in the Kraus (operator sum) representation \cite{Choi,Kraus,QCQI,Lind},%===============================================================================
\begin{equation}%                 Equation 1
\rho_{S}(t)={\textstyle \sum_{n}}\langle n|U|0\rangle\rho_{S}(0)\langle0|U^{\dagger}|n\rangle={\textstyle \sum_{n}}K_{n}\rho_{S}(0)K_{n}^{\dagger},
\label{eq:1}
\end{equation}
%===============================================================================
where $K_{n}=\langle n|U|0\rangle$ are the so-called Kraus operators,
which depend on the initial state of the environment and the choice
of the complete basis of the environment $\{|n\rangle\}$. Changing
the basis of the environment from $\{|n\rangle\}$ to $\{|m\rangle\}$
as $\langle n|=\sum_{m}\langle m|V_{n,m}$ will lead to another set
of Kraus operators $L_{m}=\langle m|U|0\rangle$, where $V$ is a
unitary matrix and the relation between the two sets of Kraus operators
can be written as %===============================================================================
\begin{equation}%                 Equation 2
K_{n}={\textstyle \sum_{m}}V_{n,m}\langle m|U|0\rangle={\textstyle \sum_{m}}V_{n,m}L_{m}.
\label{eq:2}
\end{equation}
%===============================================================================

Clearly, a measurement on the environment will collapse the environment
into an eigenstate of the measured observable. Correspondingly, the
system will also be projected into a state relative to each resultant
environmental state after measurement, i.e., $\rho_{S,n}(t)=K_n\rho_S(0)K_n^\dagger$, if we observe the $n^{\text{th}}$ outcome in the measurement of environment. Equation (\ref{eq:1}) represents
an ensemble of the system states without specifying the measurement
outcomes. If the decomposition (\ref{eq:1}) is RU, i.e., $K_{n}=c_{n}U_{n}$ for each $n$, where $U_{n}^{\dagger}=U_{n}^{-1}$
and $c_{n}$ satisfy $\sum_{n}|c_{n}|^{2}=1$, one can apply an inverse operation to recover
the initial quantum state based on the measurement outcome $n$, %===============================================================================
\begin{equation}%                 Equation 3
\rho_{R}(t)=R_{n}\rho_{S,n}(t)R_{n}^{\dagger}=\rho_{S}(0),
\label{eq:3}
\end{equation}
%===============================================================================
where $R_{n}=\frac{1}{c_{n}}U_{n}^{-1}$. This scheme of restoring
initial quantum states depends on the form of the Kraus decomposition.
If an RU-type Kraus decomposition exists, the quantum information
stored in the initial state can be fully recovered.
%                                  END of II
%*******************************************************************************
%*******************************************************************************
%                       III. $N$ Qubits in a Common Bath
\section{\label{sec:III}$N$ Qubits in a Common Bath}
Focusing on the dephasing channel \cite{YuPRB,Liu}, the simplest example,
the single-qubit case, has been analyzed in Ref.~\cite{Strunz}, where
an RU decomposition is constructed. Here, we will start from the model
of two qubits in a common bath and then extend the method to a general
$N$-qubit case. The single-qubit case can be revealed as a special case
in our framework. In the discussion in subsection \ref{sec:IV.B},
we will see the measurement required in our method is based on Fock
states, which is much simpler than the method discussed in Ref.~\cite{Strunz}. 
%-------------------------------------------------------------------------------
%       III.A. Two-Qubit Non-RU Decomposition and Entanglement Restoration
\subsection{\label{sec:III.A}Two-Qubit Non-RU Decomposition and Entanglement Restoration}
It is instructive to consider first how to recover a quantum entangled
state via a non-RU decomposition, while the restoration of arbitrary
initial states based on RU decomposition will be discussed later. For
the common bath case, the Hamiltonian of the two-qubit dephasing model is $H=\frac{\omega }{2}S_{z}+\sum_{k}\omega_{k}b_{k}^{\dagger}b_{k}+\sum_{k}g_{k}S_{z}(b_{k}+b_{k}^{\dagger})$,
where $S_{z}=\frac{1}{2}(\sigma_{z}^{A}+\sigma_{z}^{B})$. Assuming
the initial state of environment is the vacuum state $|0\rangle$, in the interaction picture,
the solution of the Schr\"{o}dinger equation is $|\psi_{tot}(t)\rangle=U|\psi_{S}(0)\rangle|0\rangle$,
where%===============================================================================
\begin{equation}%                 Equation 4
U=\prod_{k}e^{-i\phi_{k}(t)S_{z}^{2}}e^{-iS_{z}b_{k}^{\dagger}G_{k}(t)}e^{-iS_{z}b_{k}G_{k}^{*}(t)},
\label{eq:4}
\end{equation}
%===============================================================================
where $G_{k}(t)=\int_{0}^{t}g_{k}e^{i\omega_{k}s}ds$, and $\frac{d}{dt}\phi_{k}(t)=-ig_{k}e^{-i\omega_{k}t}G_{k}(t)$.
Then,%===============================================================================
\begin{equation}%                 Equation 5
\langle\{m_{k}\}|U|0\rangle=\prod_{k}e^{-i\phi_{k}(t)S_{z}^{2}}\frac{[-iS_{z}G_{k}(t)]^{m_{k}}}{m_{k}!}
\label{eq:5}
\end{equation}
%===============================================================================
where the notation $|\{m_{k}\}\rangle\equiv|m_{1},m_{2},m_{3}\cdots\rangle\equiv\otimes_{k}|m_{k}\rangle$
represents the multi-mode Fock state, where $m_{k}$ are photon numbers
in the $k^{\text{th}}$ mode. In (\ref{eq:5}), every excitation in the environment contributes one $S_{z}$ operator, so the product contains $(S_{z})^{m}$
operators ($m=\sum_{k}m_{k}$ is the total photon number). If the total
photon number $m$ is odd, we have $(S_{z})^{m}=S_{z}$; if the total
photon number $m$ is even, we have $(S_{z})^{m}=S_{z}^{2}$; if all
the $m_{k}$ are zero, only the first term $e^{-i\phi_{k}(t)S_{z}^{2}}$
is left. Finally, we have only three types of Kraus operators:

1. If $|\{m_{k}\}\rangle=|0\rangle$, then%===============================================================================
\begin{equation}%                 Equation 6
\langle\{m_{k}\}|U|0\rangle=\text{diag}\{l_{1}(t),1,1,l_{1}(t)\},
\label{eq:6}
\end{equation}
%===============================================================================
where the time-dependent element is%===============================================================================
\begin{equation}%                 Equation 7
l_{1}(t)=\exp[-\int_{0}^{t}dt^{\prime}\int_{0}^{t^{\prime}}\alpha(t^{\prime},s)ds],
\label{eq:7}
\end{equation}
%===============================================================================
with $\alpha(t^{\prime},s)=\sum_{k}g_{k}^{2}e^{-i\omega_{k}(t^{\prime}-s)}$.

2. If the total photon number $m\equiv\sum\nolimits _{k}m_{k}$ in
$|\{m_{k}\}\rangle$ is odd, then %===============================================================================
\begin{equation}%                 Equation 8
\langle\{m_{k}\}|U|0\rangle=F_{\{m_{k}\}}^{odd}(t)S_{z},
\label{eq:8}
\end{equation}
%===============================================================================
where $F_{\{m_{k}\}}^{odd}=\prod_{k}e^{-i\phi_{k}(t)}\frac{[-iG_{k}(t)]^{m_{k}}}{m_{k}!}$.

3. If the total photon number $m\equiv\sum\nolimits _{k}m_{k}$ in
$|\{m_{k}\}\rangle$ is even but not $0$, then%===============================================================================
\begin{equation}%                 Equation 9
\langle\{m_{k}\}|U|0\rangle=F_{\{m_{k}\}}^{even}(t)S_{z}^{2},
\label{eq:9}
\end{equation}
%===============================================================================
where $F_{\{m_{k}\}}^{even}=\prod_{k}e^{-i\phi_{k}(t)}\frac{[-iG_{k}(t)]^{m_{k}}}{m_{k}!}$.

The reduced density matrix can be recovered by these three types of
Kraus operators as%===============================================================================
\begin{eqnarray}%                 Equation 10
\rho_{S}(t) & = & {\textstyle \sum_{\{m_{k}\}}}\langle\{m_{k}\}|U|0\rangle\rho_{S}(0)\langle0|U^{\dagger}|\{m_{k}\}\rangle\nonumber \\
 & = & \langle0|U|0\rangle\rho_{S}(0)\langle0|U^{\dagger}|0\rangle\nonumber \\
 &  & +{\textstyle \sum_{\{m_{k}\}}^{m=odd}}\langle\{m_{k}\}|U|0\rangle\rho_{S}(0)\langle0|U^{\dagger}|\{m_{k}\}\rangle\nonumber \\
 &  & +{\textstyle \sum_{\{m_{k}\}}^{m=even}}\langle\{m_{k}\}|U|0\rangle\rho_{S}(0)\langle0|U^{\dagger}|\{m_{k}\}\rangle\nonumber \\
 & = & \langle0|U|0\rangle\rho_{S}(0)\langle0|U^{\dagger}|0\rangle\nonumber \\
 &  & +{\textstyle \sum_{\{m_{k}\}}^{m=odd}}|F_{\{m_{k}\}}^{odd}|^{2}S_{z}\rho_{S}(0)S_{z}\nonumber \\
 &  & +{\textstyle \sum_{\{m_{k}\}}^{m=even}}|F_{\{m_{k}\}}^{even}|^{2}S_{z}^{2}\rho_{S}(0)S_{z}^{2}\nonumber \\
 & = & {\textstyle \sum_{i=1}^{3}}L_{C,i}\rho_{S}(0)L_{C,i}^{\dagger}.
\label{eq:10}
\end{eqnarray}
%===============================================================================

In the matrix form, these non-RU-type Kraus operators $L_{C,i}$ can
be explicitly written as:%===============================================================================
\begin{eqnarray}%               Equations 11-13
L_{C,1} & = & \text{diag}\{l_{1}(t),1,1,l_{1}(t)\},\label{eq:11}\\
L_{C,2} & = & l_{2}(t)\text{diag}\{1,0,0,-1\},\label{eq:12}\\
L_{C,3} & = & l_{3}(t)\text{diag}\{1,0,0,1\},\label{eq:13}
\end{eqnarray}
%===============================================================================
corresponding to the vacuum state, odd state, and even state, respectively.
The time-dependent coefficients can be expressed as $l_{1}(t)=\!\exp[-\int_{0}^{t}dt^{\prime}\int_{0}^{t^{\prime}}\alpha(t^{\prime},s)ds]$,
$l_{2}(t)=\sqrt{{\textstyle \sum_{\{m_{k}\}}}|F_{\{m_{k}\}}^{odd}|^{2}}$,
and $l_{3}(t)=\sqrt{{\textstyle \sum_{\{m_{k}\}}}|F_{\{m_{k}\}}^{even}|^{2}}$.

The corresponding measurement operators for this set of Kraus operators
are $M_{C,1}=|0\rangle\langle0|$, $M_{C,2}=\sum_{\{m_{k}\}}^{m=odd}|\{m_{k}\}\rangle\langle\{m_{k}\}|$,
and $M_{C,3}=\sum_{\{m_{k}\}}^{m=even}|\{m_{k}\}\rangle\langle\{m_{k}\}|$,
where $m$ is the total photon number summed over all modes. They
satisfy $M_{C,i}M_{C,j}=\delta_{i,j}M_{C,i}$, $\sum_{i=1}^{3}M_{C,i}=I$,
and $M_{C,i}$ are all non-negative. Performing a measurement on this
set of measurement operators, the possible outcomes give the non-RU
decomposition as%===============================================================================
\begin{equation}%                Equation 14
\text{tr}_{E}[U\rho(0)U^{\dagger}M_{C,i}]=L_{C,i}\rho_{S}(0)L_{C,i}^{\dagger},\;(i=1,2,3)
\label{eq:14}
\end{equation}
%===============================================================================
with probability $p_{C,i}=\text{tr}[L_{C,i}\rho_{S}(0)L_{C,i}^{\dagger}]$.

\begin{figure}[hb]%                  FIGURE 1
\centering
\includegraphics[width=0.99\linewidth]{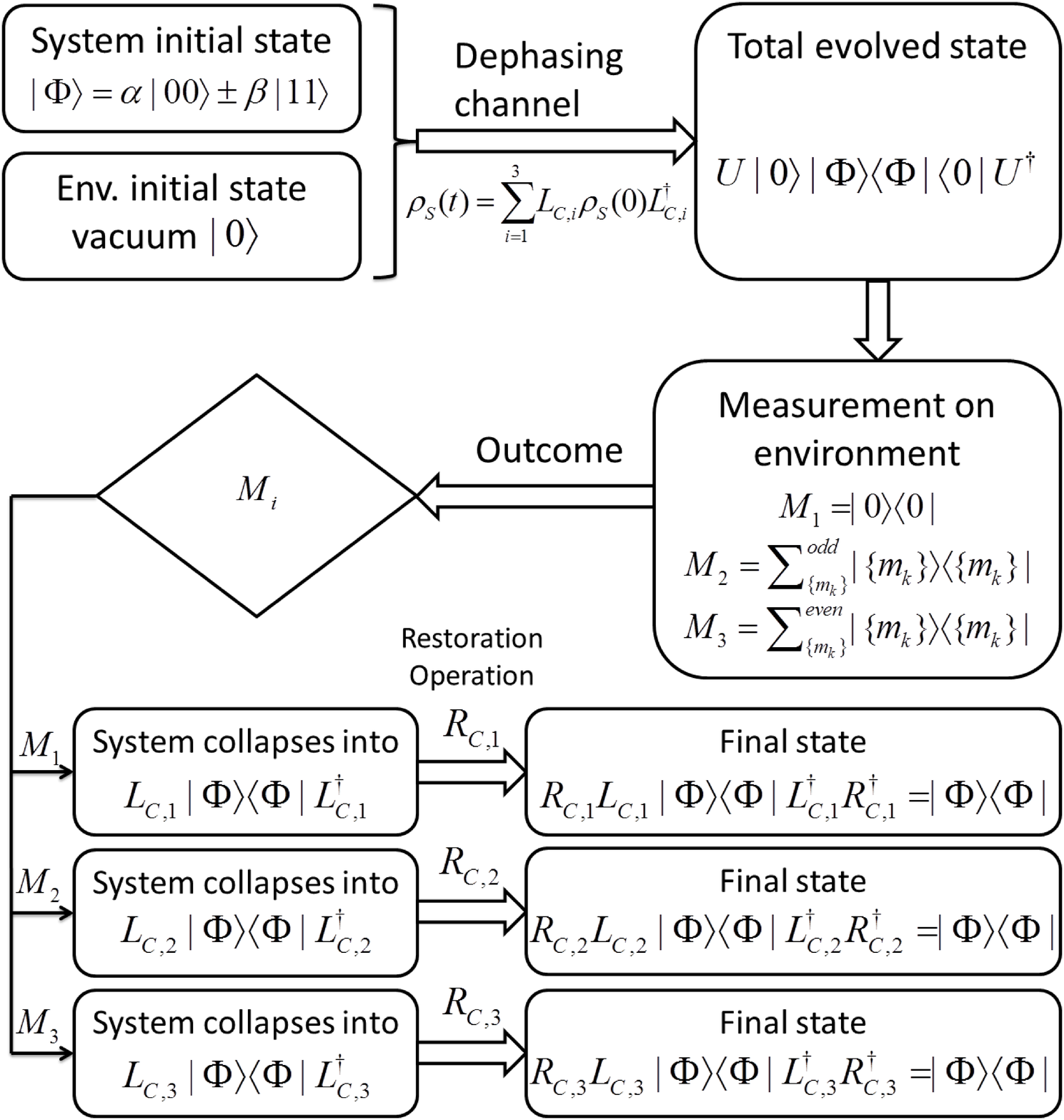}
\caption[]{\label{fig:1}Entangled states restoration scheme.}
\end{figure}

This set of Kraus operators $L_{C,i}$ are obviously non-RU-type. We show below that one can exploit the non-RU decomposition to recover
quantum entangled states. Then the restoration based on this set of
non-RU Kraus operators goes beyond the standard scheme \cite{LostFound}.
Consider two major types of initially entangled states $|\Phi\rangle=\alpha|00\rangle\pm\beta|11\rangle$
and $|\Psi\rangle=\alpha|01\rangle\pm\beta|10\rangle$, where $|\alpha|^{2}+|\beta|^{2}=1$. The second
one is decoherence-free in passage through the channel (\ref{eq:10}),
so we only focus on the first type, $|\Phi\rangle$. For this type
of initial states, the restoration operators are $R_{C,1}=l_{1}^{-1}\text{diag}\{1,1,1,1\}$,
$R_{C,2}=l_{2}^{-1}\text{diag}\{1,1,1,-1\}$, and $R_{C,3}=l_{3}^{-1}\text{diag}\{1,1,1,1\}$, which are unitary operations. It is easy to check that %===============================================================================
\begin{equation}%                Equation 15
R_{C,i}L_{C,i}|\Phi\rangle\langle\Phi|L_{C,i}^{\dagger}R_{C,i}^{\dagger}=|\Phi\rangle\langle\Phi|,\;(i=1,2,3)
\label{eq:15}
\end{equation}
%===============================================================================
meaning that the restoration operations give back the unknown initial
state precisely in all possible outcome scenarios. The experimental
setup is just measuring the environment in the Fock basis, and based
on the results $M_{C,1}$ ($|0\rangle$ state), $M_{C,2}$ (odd state),
or $M_{C,3}$ (even state), we perform the corresponding restoration
operations $R_{C,1}$, $R_{C,2}$, or $R_{C,3}$. The restoration
procedure is explicitly shown in Fig.~\ref{fig:1}.
%_______________________________________________________________________________
%_______________________________________________________________________________

Although this scheme is only applicable to some particular initial
states, the experimental setup will be relatively easy since the measurement basis is just the Fock states separated into families based on parity. Besides, in most quantum information processing schemes, entangled states like $|\Phi\rangle$ and $|\Psi\rangle$ are the most commonly used initial states. Now, we treat the case of arbitrary initial states, and then compare these two schemes.

%                                End of III.A
%-------------------------------------------------------------------------------
%-------------------------------------------------------------------------------
%                       III.B. Two-Qubit RU Decomposition
\subsection{\label{sec:III.B}Two-Qubit RU Decomposition}
In order to construct an RU decomposition and deal with arbitrary
initial states, we need to find another set of RU-type Kraus operators
which are linear combinations of the $L_{C,i}$. Consider the following
ansatz:

Suppose the RU-type Kraus operators have the form%===============================================================================
\begin{eqnarray}%              Equations 16-19
K_{C,1} & = & \sqrt{x_{1}}\text{diag}\{1,1,1,1\},\label{eq:16}\\
K_{C,2} & = & \sqrt{x_{2}}\text{diag}\{-1,1,1,1\},\label{eq:17}\\
K_{C,3} & = & \sqrt{x_{3}}\text{diag}\{1,1,1,-1\},\label{eq:18}\\
K_{C,4} & = & \sqrt{x_{4}}\text{diag}\{1,-1,-1,1\}.\label{eq:19}
\end{eqnarray}
%===============================================================================
Then, using the relation %===============================================================================
\begin{equation}%                Equation 20
\rho_{S}(t)={\textstyle \sum_{i=1}^{3}}L_{C,i}\rho_{S}(0)L_{C,i}^{\dagger}={\textstyle \sum_{j=1}^{4}}K_{C,j}\rho_{S}(0)K_{C,j}^{\dagger},
\label{eq:20}
\end{equation}
%===============================================================================
the coefficients $x_{j}$ $(j=1,2,3,4)$ are determined as
%===============================================================================
\begin{eqnarray}%               Equations 21-23
x_{1} & = & \frac{1}{4}(1+2l_{1}+|l_{3}|^{2}-|l_{2}|^{2}),\label{eq:21}\\
x_{2} & = & x_{3}=\frac{1}{4}(1-|l_{3}|^{2}+|l_{2}|^{2}),\label{eq:22}\\
x_{4} & = & \frac{1}{4}(1-2l_{1}+|l_{3}|^{2}-|l_{2}|^{2}).\label{eq:23}
\end{eqnarray}
%===============================================================================
% It is obvious that the $K_{C,i}$ are all RU-type Kraus operators, therefore
% the inverse operations showed in (\ref{eq:3}) always exist for
% each $K_{C,i}$.
Since the $K_{C,i}$ are all RU-type Kraus operators, then their inverse operations $R_{C,i}$ as in (\ref{eq:3}) always exist.

Finding the RU-type Kraus decomposition only proves the \textit{feasibility}
of perfect restoration of the initial state. In practice, we still need
to find the measurement operators corresponding to the RU-type Kraus
operators. It is well-known that any two Kraus decompositions
of the same quantum channel must be connected by a unitary matrix
\cite{QCQI,Strunz}, as in (\ref{eq:2}). The simplest Kraus decomposition
is given by $L_{\{m_{k}\}}^{\prime}=\langle\{m_{k}\}|U|0\rangle$
where we have infinite numbers of Kraus operators $L_{\{m_{k}\}}^{\prime}$
since we choose the Fock basis $\{|m_{k}\rangle\}$. Then we can follow
the scheme in \cite{Strunz} to give the measurement basis for the $K_{C,i}$.
The unitary transition matrix is just the one connecting $K_{C,i}$
and $L_{\{m_{k}\}}^{\prime}$. Although there are infinite numbers
of $L_{\{m_{k}\}}^{\prime}$, we will show that only the first several
$L_{\{m_{k}\}}^{\prime}$ operators are non-zero so that we can set
a cutoff and find the transition matrix by simply solving a set of
linear equations.

For simplicity, we use the single-mode case as an example. For a single-mode environment, the simplest $m$ Kraus operators (for a harmonic
oscillator, $m$ goes from $0$ to $\infty$), can be expressed as
$L_{m}^{\prime}=\langle m|U|0\rangle$, and the density matrix is
%===============================================================================
\begin{eqnarray}%                   Equation 24
\rho_{S}(t) & = & {\textstyle \sum_{m=0}^{\infty}}\langle m|U|0\rangle\rho_{S}(0)\langle0|U^{\dagger}|m\rangle\nonumber \\
 & = & {\textstyle \sum_{m=0}^{\infty}}L_{m}^{\prime}\rho_{S}(0)L_{m}^{\prime}{}^{\dagger}.
\label{eq:24}
\end{eqnarray}
%===============================================================================
Each Kraus operator $L_{m}^{\prime}$ can be determined by (\ref{eq:5})
or computed directly by numerical methods. Meanwhile, we have also
derived another set of RU-type Kraus operators in (\ref{eq:16}-\ref{eq:19}),
since the two sets of Kraus operators must be connected by a unitary
matrix $V$ as %===============================================================================
\begin{equation}%                   Equation 25
K_{C,n}={\textstyle \sum_{m=1}^{\infty}}V_{n,m}L_{m-1}^{\prime},
\label{eq:25}
\end{equation}
%===============================================================================
where in the case that $n>4$, $K_{C,n}$ are all zero. Then, writing
the equation in matrix form, it becomes %===============================================================================
\begin{equation}%                   Equation 26
\left[\begin{array}{c}
K_{C,n}(1,1)\\
K_{C,n}(2,2)\\
K_{C,n}(3,3)\\
K_{C,n}(4,4)
\end{array}\right]=\left[\begin{array}{ccc}
L_{0}^{\prime}(1,1) & \cdots & L_{m-1}^{\prime}(1,1)\\
L_{0}^{\prime}(2,2) & \cdots & L_{m-1}^{\prime}(2,2)\\
L_{0}^{\prime}(3,3) & \cdots & L_{m-1}^{\prime}(3,3)\\
L_{0}^{\prime}(4,4) & \cdots & L_{m-1}^{\prime}(4,4)
\end{array}\right]\!\left[\begin{array}{c}
V_{n,1}\\
V_{n,2}\\
\vdots\\
V_{n,m}
\end{array}\right]\!,
\label{eq:26}
\end{equation}
%===============================================================================
where the numbers in parentheses indicate matrix elements and the
coefficient matrix is a $4\times m$ matrix. Although the matrix is
infinite-dimensional, the probability to find a high-excitation state
$|m\rangle$ (i.e.~a very large $m$) in the environment is actually
very small. According to (\ref{eq:5}), when the total excitation $m$
is large, $L_{m}^{\prime}=\langle m|U|0\rangle$ is close to zero.
Therefore, we can take a cutoff when solving (\ref{eq:26}).
%_______________________________________________________________________________
\begin{figure}[H]%                  FIGURE 2
\centering
\includegraphics[width=0.99\linewidth]{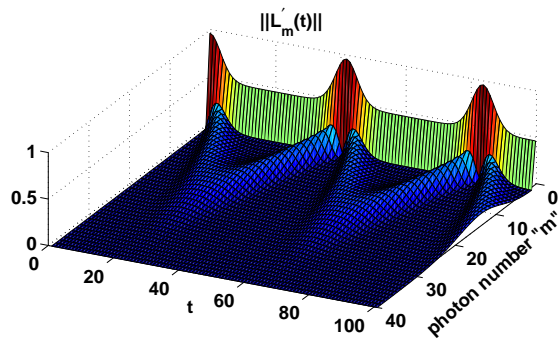}
\caption[]{\label{fig:2}(color online) $||L_{m}^{\prime}(t)||$ vs $t$ and
$m$.}
\end{figure}
%_______________________________________________________________________________

We numerically verify this in Fig.~\ref{fig:2}, in which we can
see that when the photon number is large, the trace norm ($||A||\equiv\text{tr}[\sqrt{A^{\dagger}A}]$)
of the Kraus operators $L_{m}^{\prime}(t)$ is always close to zero.
Therefore, it is safe to drop the high excitation states and the coefficient
matrix becomes finite-dimensional. For example, according to Fig.~\ref{fig:2},
we can safely choose $L_{m-1}^{\prime}=0$ when $m>30$, so then the
coefficients matrix is $4\times30$. Therefore, the equations for
$V$ form a set of first order linear equations. Since all the $L_{m-1}^{\prime}$
are known, i.e.~the coefficient matrix is fully determined, it can
be solved by computer rather quickly.

Finally, the measurement basis is just%===============================================================================
\begin{equation}%                   Equation 27
|\psi_{n}\rangle={\textstyle \sum_{m=1}^{30}}V_{n,m}^{*}|m\rangle,\quad(m,n=1,\ldots,30).\label{eq:27}
\end{equation}
%===============================================================================
The first four measurement basis states $|\psi_{n}\rangle$ ($n=1,2,3,4$)
just correspond to RU-type Kraus operators $K_{C,n}$, and the probabilities
of finding the other measurement outcomes $|\psi_{n}\rangle$ ($n>4$)
are zero \cite{Strunz}.

Although the measurement basis states for RU decomposition can be
easily numerically determined by the procedure above, the physical
meaning is still unclear since they are superpositions of Fock states,
and more importantly, it may be difficult to perform such measurements.

In the common bath case, the Fock basis does not
admit an RU decomposition. To overcome this, we have two options.
The first option is to construct an RU decomposition with Kraus operators
$K_{C,i}$, which will shift the difficulty to finding the measurement
operators. Unfortunately, such measurement operators are difficult
to realize. However, if we can overcome this by cleverly designing
an experimental measurement procedure, then the advantage of this
first choice is that the restoration scheme is applicable to arbitrary
initial states.

On the other hand, the second option is that we can make a compromise
about what we choose for the initial states. If we are only interested
in recovering some particular unknown entangled states (and in most
quantum information processing schemes, the initial states are usually
the entangled states we discussed), then we can go beyond the RU decomposition
scheme.

The non-RU decomposition may give us a much simpler way of recovering
entangled states, and simplify the implementation of the measurements. Based on our current knowledge of experimental techniques, we
show the whole scheme of the second option, and theoretically prove
the feasibility of the first option. Then, we leave the problem of
designing a clever experiment to perform a measurement on superposition
Fock states to future investigations.
%                                End of III.B
%-------------------------------------------------------------------------------
%-------------------------------------------------------------------------------
%                III.C. RU Decomposition for the $N$-Qubit Case
\subsection{\label{sec:III.C}RU Decomposition for the $N$-Qubit Case}
Using the technique of constructing RU Kraus decomposition for the two-qubit case, we can try to solve the general $N$-qubit case. Therefore, to gain some insight, we will explicitly construct the RU decomposition of the three-qubit case as a special example. First, we will review some properties of the dephasing channel which have been studied in \cite{Jing-Yu}. According to \cite{Jing-Yu}, the $N$-qubit dephasing channel can be expressed as %===============================================================================
\begin{equation}%                   Equation 28
\rho_{S}(t)=C(N)\circ\rho_{S}(0),
\label{eq:28}
\end{equation}
%===============================================================================
where $A\circ B$ is the entry-wise product of matrices $A$ and $B$
(also called a Schur or Hadamard product). Although the dimension
of matrix $C(N)$ is $N^{2}$, the rank $C(N)$ is
%===============================================================================
\begin{equation}%                   Equation 29
R(N)=\text{rank}[C(N)]=N+1,
\label{eq:29}
\end{equation}
%===============================================================================
i.e.~only $R(N)$ rows or columns give us useful information. Therefore
we can always shift the linearly independent rows (columns) to the
top-left of the whole matrix, in which case it takes the form, %===============================================================================
\begin{equation}%                   Equation 30
C(N)\Rightarrow\left[\begin{array}{cc}
M & M^{\prime}\\
M^{\prime} & M^{\prime\prime}
\end{array}\right],
\label{eq:30}
\end{equation}
%===============================================================================
where $M$ is the linearly independent part ($\text{dim}(M)\!=\!\text{rank}[C(N)]\!=\! N+1$)
while $M^{\prime}$ and $M^{\prime\prime}$ are linearly dependent
parts. Moreover, the matrix elements of $C(N)$ only contain $\{\gamma^{0}(t),\gamma^{1^{2}}(t),\gamma^{2^{2}}(t),\cdots,\gamma^{(N-1)^{2}}(t),\gamma^{N^{2}}(t)\}$,
where $\gamma(t)$ describes the decay of off-diagonal elements. The
elements in $M$ are $\gamma^{N^{2}}$ to $\gamma^{0}$, sweeping in
diagonal groups anti-diagonally from the top-right to the main-diagonal elements \cite{Jing-Yu}, so that %===============================================================================
\begin{equation}%                   Equation 31
M=\left[\begin{array}{cccc}
1 & \cdots & \gamma^{(N-1)^{2}} & \gamma^{N^{2}}\\
\vdots & 1 & \cdots & \gamma^{(N-1)^{2}}\\
\gamma^{(N-1)^{2}} & \cdots & 1 & \vdots\\
\gamma^{N^{2}} & \gamma^{(N-1)^{2}} & \cdots & 1
\end{array}\right].
\label{eq:31}
\end{equation} %===============================================================================

Given these facts, the following procedure shows how to construct the RU decomposition for an arbitrary $N$-qubit dephasing channel. 

First, write down the Schur matrix $C(N)$ of the channel, and delete
the rows (columns) which are not linearly independent. Then, we can
choose $N_{K}$ (we will give this number below) basis operators $\{B_{1},\cdots,B_{N_{K}}\}$
which are all diagonal matrices in which the diagonal elements are
either $1$ or $-1$, e.g., %===============================================================================
\begin{equation}%                   Equation 32
B_{1}=\text{diag}\{1,1,\cdots,1\},\quad B_{N_{K}}=\text{diag}\{1,-1,\cdots,-1\}.
\label{eq:32}
\end{equation}
%===============================================================================
Given this basis, the RU-type Kraus operators are %===============================================================================
\begin{equation}%                   Equation 33
K_{i}=\sqrt{c_{i}}B_{i}\quad(i=1,\cdots,N_{K}),
\label{eq:33}
\end{equation}
%===============================================================================
where $c_{i}$ are coefficients which can be determined by %===============================================================================
\begin{equation}%                   Equation 34
{\textstyle \sum_{i=1}^{N_{K}}}K_{i}\rho_{S}(0)K_{i}^{\dagger}=C(N)\circ\rho_{S}(0).
\label{eq:34}
\end{equation}
%===============================================================================

Next, determine how many (i.e., the number $N_{K}$) basis operators
are needed for the RU decomposition. In $C(N)$, only $M$ is useful while $M^{\prime}$ and $M^{\prime\prime}$
contain no new information. Thus, the number of independent equations in (\ref{eq:34})
is just the number of the upper triangular elements in $M$ plus one,
i.e.,
%===============================================================================
\begin{equation}%                   Equation 35
N_{K}=1+{\textstyle \sum_{k=1}^{N}}k=1+\frac{N(N+1)}{2},
\label{eq:35}
\end{equation}
%===============================================================================
since $M$ is $(N+1)\times(N+1)$. That is the reason we need to choose
$N_{K}$ basis elements; we need $N_{K}$ coefficients $c_{i}$ to
satisfy those equations given by (\ref{eq:34}). Solving for the
coefficients $c_{i}$ $(i=1,\cdots,N_{K})$ and substituting them
back into (\ref{eq:33}), all the Kraus operators for the RU decomposition
of our $N$-qubit dephasing channel are fully determined. There are
a total of $2^{2^{N}}$ choices of basis, but we only need $N_{K}=1+\frac{N(N+1)}{2}$
basis elements, so the choice of basis is not unique.

It will be helpful to understand the above scheme of constructing
the RU decomposition of the $N$-qubit model by explicitly solving the three-qubit case as an example. In the three-qubit case, the matrix $C(3)$ is
%===============================================================================
\begin{equation}%                   Equation 36
C(3)=\left[\begin{array}{cccccccc}
1 & \gamma & \gamma & \gamma^{4} & \gamma & \gamma^{4} & \gamma^{4} & \gamma^{9}\\
\gamma & 1 & 1 & \gamma & 1 & \gamma & \gamma & \gamma^{4}\\
\gamma & 1 & 1 & \gamma & 1 & \gamma & \gamma & \gamma^{4}\\
\gamma^{4} & \gamma & \gamma & 1 & \gamma & 1 & 1 & \gamma\\
\gamma & 1 & 1 & \gamma & 1 & \gamma & \gamma & \gamma^{4}\\
\gamma^{4} & \gamma & \gamma & 1 & \gamma & 1 & 1 & \gamma\\
\gamma^{4} & \gamma & \gamma & 1 & \gamma & 1 & 1 & \gamma\\
\gamma^{9} & \gamma^{4} & \gamma^{4} & \gamma & \gamma^{4} & \gamma & \gamma & 1
\end{array}\right].
\label{eq:36}
\end{equation}
%===============================================================================
The rank of $C(3)$ is $R(3)=4$, so we need at least $1+\frac{3(3+1)}{2}=7$
independent Kraus operators for the RU decomposition. The
choice of Kraus operators is not unique, and one example is %===============================================================================
\begin{eqnarray}%                Equations 37-43
K_{1} & = & \sqrt{A}\text{diag}\{1,1,1,1,1,1,1,1)\},\label{eq:37}\\
K_{2} & = & \sqrt{B}\text{diag}\{-1,1,1,1,1,1,1,1\},\label{eq:38}\\
K_{3} & = & \sqrt{C}\text{diag}\{1,-1,-1,1,-1,1,1,1\},\label{eq:39}\\
K_{4} & = & \sqrt{D}\text{diag}\{1,1,1,-1,1,-1,-1,1\},\label{eq:40}\\
K_{5} & = & \sqrt{E}\text{diag}\{1,1,1,1,1,1,1,-1\},\label{eq:41}\\
K_{6} & = & \sqrt{F}\text{diag}\{-1,-1,-1,1,-1,1,1,1\},\label{eq:42}\\
K_{7} & = & \sqrt{G}\text{diag}\{1,-1,-1,-1,-1,-1,-1,1\}.\label{eq:43}
\end{eqnarray}
%===============================================================================
Then, according to the relation %===============================================================================
\begin{equation}%                  Equation 44
\rho_{S}(t)=C(3)\circ\rho_{S}(0)={\textstyle \sum_{i=1}^{7}}K_{i}\rho_{S}(0)K_{i}^{\dagger},
\label{eq:44}
\end{equation}
%===============================================================================
we can obtain the matrix equation %===============================================================================
\begin{equation}%                  Equation 45
\left[\begin{array}{rrrrrrr}
1 & 1 & 1 & 1 & 1 & 1 & 1\\
1 & -1 & -1 & 1 & 1 & 1 & -1\\
1 & 1 & -1 & -1 & 1 & -1 & 1\\
1 & 1 & 1 & -1 & -1 & 1 & -1\\
1 & -1 & 1 & -1 & 1 & -1 & -1\\
1 & 1 & -1 & 1 & -1 & -1 & -1\\
1 & -1 & 1 & 1 & -1 & -1 & 1
\end{array}\right]\left[\begin{array}{c}
A\\
B\\
C\\
D\\
E\\
F\\
G
\end{array}\right]=\left[\begin{array}{c}
1\\
\gamma\\
\gamma\\
\gamma\\
\gamma^{4}\\
\gamma^{4}\\
\gamma^{9}
\end{array}\right].
\label{eq:45}
\end{equation}
%===============================================================================
Therefore, the solution is %===============================================================================
\begin{equation}%                  Equation 46
\left[\begin{array}{c}
A\\
B\\
C\\
D\\
E\\
F\\
G
\end{array}\right]=M_{coef}^{-1}\left[\begin{array}{c}
1\\
\gamma\\
\gamma\\
\gamma\\
\gamma^{4}\\
\gamma^{4}\\
\gamma^{9}
\end{array}\right],
\label{eq:46}
\end{equation}
%===============================================================================
where $M_{coef}$ is the coefficient matrix in (\ref{eq:45}).
%                                End of III.C
%-------------------------------------------------------------------------------
%                                 END of III
%*******************************************************************************
%*******************************************************************************
%                       IV. $N$ Qubits in Individual Baths
\section{\label{sec:IV}$N$ Qubits in Individual Baths}
Another type of system-environment interaction for the $N$-qubit model
is $N$ qubits interacting with individual baths. In order to put
it into perspective, we will first consider the one-qubit dephasing
model. Although the RU decomposition for one qubit is already
discussed in \cite{Strunz}, we still want to show an alternative
set of Kraus operators which corresponds to a better measurement basis. This will solve the measurement difficulty in Ref.~\cite{Strunz},
where for a more complex environment it is not clear how to find the
environment measurement basis that realizes their RU decomposition.
Then, our RU decomposition for the one-qubit system will be easily
generalized to the $N$-qubit case. Now, we will derive the
RU-type Kraus decomposition for a general $N$-qubit system based
on an explicit measurement basis where a physical implementation of
the scheme is apparent.
%-------------------------------------------------------------------------------
%   IV.A. Parity Measurement of the Cavity and the One-Qubit RU Decomposition
\subsection{\label{sec:IV.A}Parity Measurement of the Cavity and the One-Qubit RU Decomposition}
The total quantum system is described by the Hamiltonian
$H=\frac{\omega}{2}\sigma_{z}+\sum_{k}\omega_{k}b_{k}^{\dagger}b_{k}+\sum_{k}g_{k}\sigma_{z}(b_{k}+b_{k}^{\dagger})$,
which represents a single qubit interacting with a set of harmonic
oscillators. In the interaction picture, the Hamiltonian becomes %===============================================================================
\begin{equation}%                  Equation 47
H_{I}={\textstyle \sum_{k}}g_{k}\sigma_{z}(b_{k}e^{-i\omega_{k}t}+b_{k}^{\dagger}e^{i\omega_{k}t}),
\label{eq:47}
\end{equation}
%===============================================================================
Comparing this to the $2$-qubit case discussed in Sec.~\ref{sec:III.A},
we can replace the operator $S_{z}$ by $\sigma_{z}$ to obtain
%===============================================================================
\begin{equation}%                  Equation 48
\langle\{m_{k}\}|U|0\rangle=\prod_{k}e^{-i\phi_{k}(t)}\frac{[-i\sigma_{z}G_{k}(t)]^{m_{k}}}{m_{k}!}.
\label{eq:48}
\end{equation}
%===============================================================================
In contrast to the behavior of $S_{z}$, we have $\sigma_{z}^{2}=\sigma_{z}^{0}=I$,
which compresses the three cases into two cases. The two types of Kraus
operators are given by

1. If the total photon number $m\equiv\sum\nolimits _{k}m_{k}$ in
$|\{m_{k}\}\rangle$ is odd, then %===============================================================================
\begin{equation}%                  Equation 49
\langle\{m_{k}\}|U|0\rangle=F_{\{m_{k}\}}^{odd}(t)\sigma_{z},
\label{eq:49}
\end{equation}
%===============================================================================
where $F_{\{m_{k}\}}^{odd}=\prod_{k}e^{-i\phi_{k}(t)}\frac{[-iG_{k}(t)]^{m_{k}}}{m_{k}!}$.

2. If the total photon number $m\equiv\sum\nolimits _{k}m_{k}$ in
$|\{m_{k}\}\rangle$ is even, then %===============================================================================
\begin{equation}%                  Equation 50
\langle\{m_{k}\}|U|0\rangle=F_{\{m_{k}\}}^{even}(t)I,
\label{eq:50}
\end{equation}
%===============================================================================
where $F_{\{m_{k}\}}^{even}=\prod_{k}e^{-i\phi_{k}(t)}\frac{[-iG_{k}(t)]^{m_{k}}}{m_{k}!}$.
The vacuum state case that appeared in the two-qubit case merged into this even-state case.

Finally, the Kraus decomposition can be written as
%===============================================================================
\begin{equation}%                  Equation 51
\rho_{S}(t)=K_{1}\rho_{S}(0)K_{1}^{\dagger}+K_{2}\rho_{S}(0)K_{2}^{\dagger},\label{eq:51}
\end{equation}
%===============================================================================
where $K_{1}=\sqrt{{\textstyle \sum_{\{m_{k}\}}}|F_{\{m_{k}\}}^{odd}|^{2}}\sigma_{z}$,
and $K_{2}=\sqrt{{\textstyle \sum_{\{m_{k}\}}}|F_{\{m_{k}\}}^{even}|^{2}}I$.
Then, our two Kraus operators are explicitly given with the basis.
Since these Kraus operators are already in RU form, we can directly use them to recover arbitrary initial states.  Also,
the physical meaning of this set of RU-type Kraus operators is
clear. If a measurement of the photon (excitation) numbers in the
environment is performed, then two possible outcomes, odd or even,
will correspond to the RU-type Kraus operators $\sigma_{z}$ and $I$,
respectively.

Now, considering a positive operator-valued measure (POVM) described
by the measurement operators, %===============================================================================
\begin{eqnarray}%                Equations 52-53
M_{1} & = & {\textstyle \sum_{\{m_{k}\}}^{m=odd}|\{m_{k}\}\rangle\langle\{m_{k}\}|,}\label{eq:52}\\
M_{2} & = & {\textstyle \sum_{\{m_{k}\}}^{m=even}|\{m_{k}\}\rangle\langle\{m_{k}\}|,}\label{eq:53}
\end{eqnarray}
%===============================================================================
it is easy to check they satisfy $M_{i}M_{j}=\delta_{i,j}M_{i}$,
$M_{1}+M_{2}=I$, and that they are non-negative. Each measurement operator
gives one specific realization as $\text{tr}_{E}[U\rho(0)U^{\dagger}M_{i}]=K_{i}\rho_{S}(0)K_{i}^{\dagger}\quad(i=1,2)$.

The measurement required is just a parity measurement. For example,
if an odd photon number is detected, one immediately concludes that
the system must have collapsed into the state $\rho_{S,1}(t)=K_{1}\rho_{S}(0)K_{1}^{\dagger}$.
Since $K_{1}$ and $K_{2}$ are each proportional to a unitary matrix, it is easy to see that an inverse operation can fully restore the initial
state as shown in (\ref{eq:3}) with $R_{1}=\sigma_{z}/\sqrt{\sum_{\{m_{k}\}}|F_{\{m_{k}\}}^{odd}|^{2}}$
and $R_{2}=I/\sqrt{\sum_{\{m_{k}\}}|F_{\{m_{k}\}}^{even}|^{2}}$.
The restoration procedure is similar to the two-qubit case plotted
in Fig.~\ref{fig:1}.

It should be noted that \cite{Strunz} also finds another RU decomposition
of this model, and they have shown how to derive the measurement operators.
In their paper, the measurement basis is time-dependent, while in our scheme,
the measurement basis is time-independent. The purpose of using this 
measurement basis (parity of Fock states) is to find an explicit experiment realization 
of the scheme \cite{You,parity1,parity2}.
%                                End of IV.A
%-------------------------------------------------------------------------------
%-------------------------------------------------------------------------------
%                     IV.B. $N$ Qubits in Individual Baths
\subsection{\label{sec:IV.B}$N$ Qubits in Individual Baths}
Based on the discussion of the restoration scheme for the single-qubit
model, we can extend the scheme to $N$ qubits in the case of individual
baths. Consider the Hamiltonian $H=\sum_{k=1}^{N}\frac{\omega_{k}}{2}\sigma_{z}^{k}+\sum_{k=1}^{N}\sum_{l}\Omega_{k,l}b_{k,l}^{\dagger}b_{k,l}+\sum_{k=1}^{N}\sum_{l}g_{k,l}\sigma_{z}^{k}(b_{k,l}+b_{k,l}^{\dagger})$.
In the case of individual baths, it is easy to see that the evolution
of the total system can be written as a tensor product of all the
participating subsystems. The Kraus operators for the total can be
obtained by employing the tensor products of the Kraus operators for
each subsystem, i.e.,
%===============================================================================
\begin{equation}%                 Equation 54
K_{S,n}=K_{1,j_{1}}\otimes K_{2,j_{2}}\otimes K_{3,j_{3}}\otimes\cdots\otimes K_{N,j_{N}},
\label{eq:54}
\end{equation}
%===============================================================================
where $K_{S,n}$ ($n=1,\ldots,2^{N}$) are the Kraus operators for
the $N$-qubit system in the case of individual baths, and $K_{i,j_{i}}$
are the $j_{i}^{\text{th}}$ Kraus operators of the $i^{\text{th}}$
subsystem. As discussed in the last section, there are two Kraus operators
for each subsystem, so the index $j_{i}$ can be $j_{i}=1$ or $j_{i}=2$
for each subsystem from $1$ to $N$. The total number of Kraus operators
of the $N$-qubit system is $2^{N}$, so the index $n$ ranges from
$1$ to $2^{N}$.

For simplicity, but without losing generality, we give an explicit
example for the special case of $N=2$, i.e.~the two-qubit model.
In this case, the two-qubit system can be described by %===============================================================================
\begin{equation}%                 Equation 55
\rho(t)={\textstyle \sum_{n=1}^{4}}K_{S,n}\rho(0)K_{S,n}^{\dagger}.
\label{eq:55}
\end{equation}
%===============================================================================
The four Kraus operators can be explicitly written as tensor products
of the Kraus operators of the single qubit case; $K_{S,1}=K_{1,1}\otimes K_{2,1}$,
$K_{S,2}=K_{1,1}\otimes K_{2,2}$, $K_{S,3}=K_{1,2}\otimes K_{2,1}$,
and $K_{S,4}=K_{1,2}\otimes K_{2,2}$. The measurement basis for each
subsystem has already been analyzed in the last section. In the two-qubit
case, we need to measure both environment ``1'' and ``2'', and
each measurement could return one possible outcome, ``odd'' or ``even.''

Finally, the four total system outcomes correspond to four Kraus operators,
each belonging to one possible final state of the system $K_{S,n}\rho(0)K_{S,n}^{\dagger}$.
Since the Kraus operators for each subsystem are RU-type, the total
Kraus decomposition is also RU-type, which allows us to find an inverse
operation to fully restore the initial state.
%                                End of IV.B
%-------------------------------------------------------------------------------
%                                 END of IV
%*******************************************************************************
%*******************************************************************************
%                              V. Conclusions
\section{\label{sec:V}Conclusions}
In this paper, we proposed an environment-assisted error correction scheme to eliminate the quantum error
caused by a dephasing channel. The correction scheme can be accomplished
with one projective measurement on the environment and one unitary reversal operation on the system. The restoration procedure is deterministic,
not probabilistic, and in principle the success probability is always
$1$. We showed that the required measurement on the environment
can be performed in the Fock basis, requiring only the parity of the photon numbers in all the participating modes.

In addition, we showed that the dephasing error can be eliminated from
a multiple-qubit system by explicitly constructing the RU decomposition
for the $N$-qubit case. In the example of a two-qubit system coupled
to a common bath, we went beyond the original restoration scheme proposed
in \cite{LostFound}, which is based on RU decomposition. We showed that some non-RU decompositions can also be used to recover particular types of entangled states. This may open a new path to find more environment-assisted correction schemes based on RU or non-RU decomposition of quantum channels.
%                                  END of V
%*******************************************************************************
%*******************************************************************************
%                               ACKNOWLEDGEMENTS
\begin{acknowledgments}
We thank J. H. Eberly and W. Strunz for useful discussions. We acknowledge grant
support from the NSF PHY-0925174, DOD/AF/AFOSR No. FA9550-12-1-0001.
\end{acknowledgments}
%                            END of ACKNOWLEDGEMENTS
%*******************************************************************************
%*******************************************************************************
%                                 BIBLIOGRAPHY

%                             END of BIBLIOGRAPHY
%*******************************************************************************
\end{document}